\documentclass[sigconf]{acmart}

\AtBeginDocument{%
  \providecommand\BibTeX{{%
    \normalfont B\kern-0.5em{\scshape i\kern-0.25em b}\kern-0.8em\TeX}}}

\setcopyright{acmcopyright}
\copyrightyear{2018}
\acmYear{2018}
\acmDOI{10.1145/1122445.1122456}

\acmConference[Woodstock '18]{Woodstock '18: ACM Symposium on Neural
  Gaze Detection}{June 03--05, 2018}{Woodstock, NY}
\acmBooktitle{Woodstock '18: ACM Symposium on Neural Gaze Detection,
  June 03--05, 2018, Woodstock, NY}
\acmPrice{15.00}
\acmISBN{978-1-4503-XXXX-X/18/06}

\usepackage{xcolor}
\usepackage{multirow}
\usepackage{graphicx}
\usepackage{subcaption}
\usepackage{caption}
\usepackage{enumitem}
\usepackage[colorinlistoftodos]{todonotes}

\usepackage{makecell}

\begin{document}

\title{Multi-Interest-Aware User Modeling for Large-Scale Sequential Recommendations}

\author{Jianxun Lian } 
\email{jianxun.lian@microsoft.com}  
\affiliation{%
  \institution{Microsoft Research Asia }
  \state{Beijing}
  \country{China}
}
\author{Iyad Batal } 
\email{iybatal@microsoft.com}  
\affiliation{%
  \institution{Microsoft Bing Ads} 
  \city{Sunnyvale}
  \state{California}
  \country{United States}
}

\author{Zheng Liu} 
\email{zheng.liu@microsoft.com}  
\affiliation{%
  \institution{Microsoft Research Asia } 
  \state{Beijing}
  \country{China}
}

\author{Akshay Soni} 
\email{akson@microsoft.com}  
\affiliation{%
  \institution{Microsoft Bing Ads} 
  \city{Sunnyvale}
  \state{California}
  \country{United States}
}

\author{Eun Yong Kang} 
\author{Yajun Wang} 
\email{{eun.kang,yajunw}@microsoft.com}  
\affiliation{%
  \institution{Microsoft Bing Ads} 
  \city{Sunnyvale}
  \state{California}
  \country{United States}
}

\author{Xing Xie} 
\email{xing.xie@microsoft.com}  
\affiliation{%
  \institution{Microsoft Research Asia } 
  \state{Beijing}
  \country{China}
}
 
\renewcommand{\shortauthors}{Jianxun Lian, Iyad Batal, et al.}

\begin{abstract}
Precise user modeling is critical for online personalized recommendation services. Generally, users' interests are diverse and are not limited to a single aspect, which is particularly evident when their behaviors are observed for a longer time. For example, a user may demonstrate interests in cats/dogs, dancing and food \& delights when browsing short videos on Tik Tok; the same user may show interests in real estate and women's wear in her web browsing behaviors. Traditional models tend to encode a user's behaviors into a single embedding vector, which do not have enough capacity to effectively capture her diverse interests.

This paper proposes a Sequential User Matrix (SUM) to accurately and efficiently capture users' diverse interests. SUM models user behavior with a multi-channel network, with each channel representing a different aspect of the user's interests. User states in different channels are updated by an \emph{erase-and-add} paradigm with interest- and instance-level attention. We further propose a local proximity debuff component and a highway connection component to make the model more robust and accurate. SUM can be maintained and updated incrementally, making it feasible to be deployed for large-scale  online serving. We conduct extensive experiments on two datasets. Results demonstrate that SUM consistently outperforms state-of-the-art baselines.
\end{abstract}
\settopmatter{printacmref=false} 

\ccsdesc[500]{Information systems~Computational advertising}
\ccsdesc[500]{Information systems~Collaborative filtering}
\ccsdesc[500]{Information systems~Personalization}
\ccsdesc[500]{Information systems~Recommender systems}

\keywords{sequential recommendation, multiple interests, memory networks}

\maketitle

\section{Introduction}
Sequential recommender systems have attracted a lot of attention in both academia \cite{chen2018sequential,hidasi2018recurrent,yu2019adaptive,10.1145/3292500.3330984,10.1145/3308558.3313408,10.1145/3366423.3380190} and industry \cite{zhou2019deep,pi2019practice,10.1145/3097983.3098108,10.1145/3357384.3357818} in recent years. Different from general recommender systems \cite{koren2009matrix,koren2008factorization} that aim to learn users' long-term preference, sequential recommender systems take the sequence of user behaviors as context and predict her short-term interests, such as what items she will interact with in the near future, or to the extreme case, what items she will interact with next. Sequential recommendation models can capture users' dynamic and evolving interests over time. In the past few years, a lot of related models have been proposed for sequential user modeling, including Recurrent Neural Network (RNN) based models \cite{hidasi2015session,hidasi2018recurrent,10.1145/3097983.3098108,yu2019adaptive,zhou2019deep,li2017neural}, Convolutional Neural Network (CNN) based models \cite{10.1145/3289600.3290975,tang2018personalized}, and Self-attention based models \cite{DBLP:conf/icdm/KangM18,10.1145/3357384.3357895,DBLP:journals/corr/abs-1808-06414}. Although state-of-the-art results have been reported by these approaches, most of them cannot be applied in large scale online systems due to the tight latency requirement in real-time online serving. A typical real-time recommendation service requires the model to respond with results in less than ten milliseconds. Given such a constraint, how to effectively model users' sequential behaviors, especially when the behavior sequence is long, becomes an crucial and challenging task.

Gated Recurrent Unit (GRU) \cite{cho-etal-2014-properties}, despite the simplicity of its structure, turns out to be effective in sequential modeling and is widely deployed in industry \cite{10.1145/3097983.3098108,zhou2019deep}. There are two main merits of GRU. First, it contains a gated mechanism to control how information is passed through or forgotten, the gradient vanishing and exploding problems are alleviated so that GRU cell can handle better relatively longer sequences compared with vanilla RNN cell. Second, GRU-based models support inference in an \textbf{incremental update} manner, which makes it practical for online serving. Incremental update means that there is no need to store and process the whole user behavior sequence all over again at each time when the recommender system receives a new user event. Instead, we just need to maintain a user state vector for every user; when a new event comes in, we update the user state only based on this single event so that the model can respond quickly. After we deployed a GRU-based recommender system to replace our previous attention-based neural model (which is non-sequential), the online click-through rate (CTR) was significantly improved by 0.82\% to 4.54\% across different scenarios in Bing Native Advertising business. 

However, GRU only generates a single embedding vector to represent a user state; when user behavior sequence becomes longer, she may reveal multiple interests that belong to different topics and are not suitable to be clustered into one representation. For example, for short video recommendation, a user may continuously browse a dozen of short videos on Tik Tok. The browsed videos may include different topics like Cute Pets,  Food \& Delights and Health \& Fitness; in online advertising scenarios, a user may browse web pages related to used cars, men shoes and kids furniture. Thus, a single user vector heavily restricts the representation ability of the model, and this defect cannot be remedied simply by increasing the embedding size of GRU. 
In this paper, we propose Sequential User Matrix (SUM), which is built on but substantially improves a classical Recommender system with User Memory network~(RUM) \cite{chen2018sequential} for better modeling of users' multiple interests in a sequence. There are mainly three novel mechanisms proposed in SUM. \\
\emph{Interest-level and instance-level attention}: The writing operations in RUM generates a channel-wise attention score for each input event to indicate the event's relatedness to different channels (aka interest level). However, within each channel, the \textsl{erase-and-add} update strategy only depends on the input event itself, without considering the instance-level context. We argue that since different user events have different importance scores, distinguishing events inside the same channel is critical for improving the model expressiveness. Thus, we propose to manipulate memory networks at both interest and instance level (Refer to Section \ref{sec:attention}).\\
\emph{Local proximity debuff}: In many applications, user behaviors tend to have the local proximity property. For example, a user's in-session web-page browsing behaviors are usually very similar; in online shopping scenario, a user's consecutive behaviors may belong to the same purchase intent. Motivated by this, we propose \textsl{a debuff mechanism} for adjacent similar behaviors, which further improves the model's capacity on handling long sequences compared with GRU. (Refer to Section \ref{sec:lpd}) \\
\emph{Highway channel}: To make a memory network model capable to recognize the exact order of behavior sequence and automatically balance the interest disentanglement and interest mixture, we reserve a channel in SUM's memory network to be \textsl{a highway connection channel}, so that every user behavior will interact with this highway channel without being influenced by the interest-level attention score. We further update the reading operation to make it attend to the states better (Refer to Section \ref{sec:hw} and \ref{sec:readingop}).

We conduct extensive offline experiments on two real-world datasets and online A/B test experiments at Bing Native Advertising scenario. Experimental results demonstrate that SUM outperforms competitive baselines significantly and consistently. In addition, we provide some case studies to verify that SUM indeed captures a user's diverse interests in its different memory channels. We open-source SUM at \url{https://aka.ms/ms-sum}.

\begin{figure*}[t]
  \centering
  \includegraphics[trim=0 0 0 0,clip,width=1.0\linewidth]{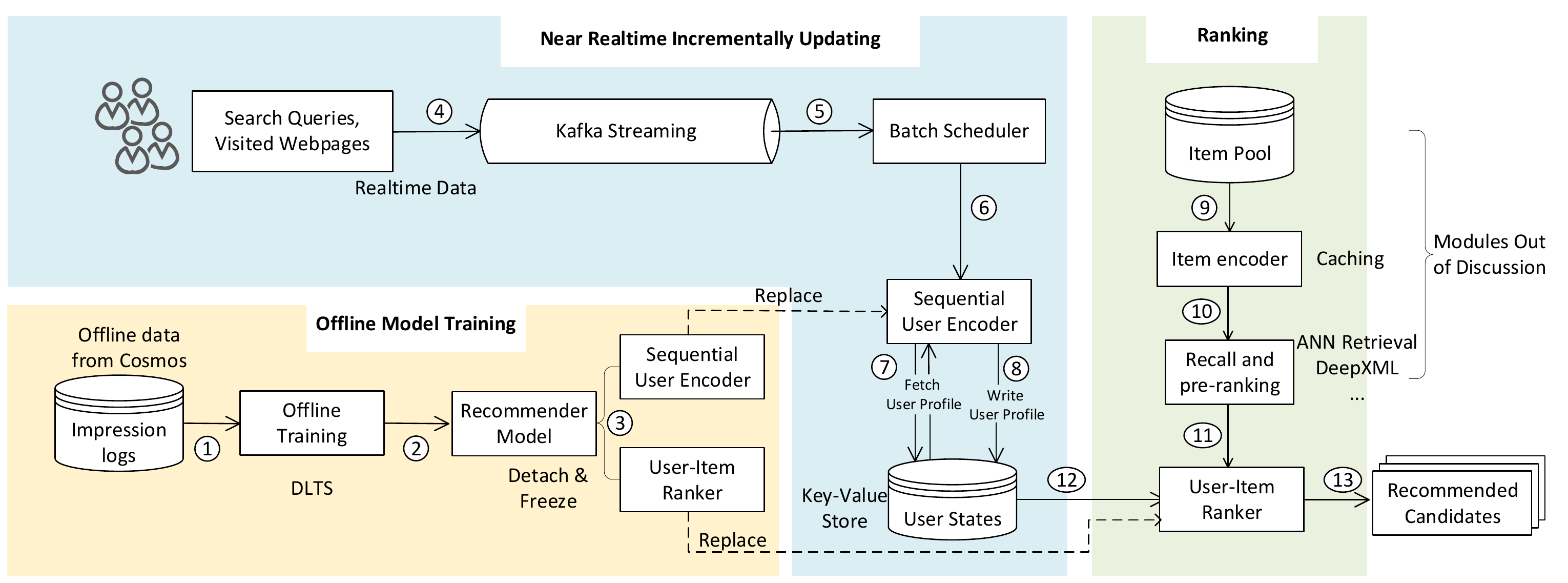}
  \caption{An overview of our near real-time (NRT) recommendation serving system. Numbers are simply for better understanding purpose, not necessarily meaning the exact order of execution.}
\label{fig:system}
\end{figure*}

\section{Near Real-time Recommender System at Bing Advertising}
\label{sec:nrt}
We first describe the Bing Native Advertising scenario and our near real-time~(NRT) system for large-scale online serving. We display personalized ads to users in a native manner \footnote{\url{https://en.wikipedia.org/wiki/Native_advertising}} when users connect to our services, such as browsing Microsoft News. Because ads clicking behaviors are extremely sparse, we choose to use massive web behaviors, including users' visited pages, Bing search queries and search clicks for more comprehensive user modeling. Through offline experiments, we find that sequential models are much better than non-sequential models because the former can automatically learn the sequential signal  and capture the temporal evolution in users' interests. However, deploying a sequential model for large-scale online serving is challenging, due to the fact that processing users' long behavior sequences in real-time is expensive. To address this issue, a popular design for large-scale recommendation systems is to decouple the architecture into two components \cite{pi2019practice}: the sequential user modeling component (aka \textsl{user encoder}) and the user-item preference scorer (aka \textsl{ranker}). These two components are running on two separate applications. The user encoder maintains the latest embeddings for users. Once a new user behavior happens, it updates the user's embedding incrementally rather than re-compute the whole behavior sequence from the beginning. This module is usually built as the near realtime module. On the other hand, the user-item preference scorer reads the latest user embeddings directly from memory so that the latency of system response is minimized.

Following this decoupled paradigm, we build an NRT serving system depicted in Figure \ref{fig:system}. The NRT system aims at serving any sequential models which can be updated incrementally, such as GRU. There are three pipelines in Figure \ref{fig:system}, covered by different background colors. Model training happens periodically on an offline platform called Deep Learning Training Service (DLTS). After the model training finishes, we detach the two most important parts, i.e., the sequential user encoder and the user-item ranker, from the model and freeze their parameters for serving. The blue pipeline is for sequential user state updating. Once a user performs an activity (e.g., she enters a Bing search), the event will be sent to the real-time feedback data pipeline for batch updating. For each update unit, the system first fetches the current user state from a distributed in-memory key-value store which we called \textsl{object storage}, then conducts a one-step inference based on the current user event and user state, and finally writes the new user state to the object storage. Under this incrementally updating paradigm, the NRT system can efficiently model users' interests from an extremely long sequence. As for online ranking, the system will fetch a user state from object storage directly, then concatenate them with item vectors and run the ranker module. In this paper, we focus on the user modeling module. Some other parts such as recall and pre-ranking (we use multiple techniques such as ANN \cite{liu2005investigation} and DeepXML \cite{deepxml}) are out of the discussion scope. So far, our NRT system serves hundreds of millions of unique user ids per day, with a peak QPS (queries per second) reaching 120k and the corresponding updating latency surprisingly being less than 1 minute. Thus, users' latest states can be updated in a near real-time manner for downstream applications such as personalized ranking. 

In the next section, we formally describe how we implement sequential user matrix~(SUM), which is a multi-interest-aware user model, on the NRT system.

\begin{figure}[t]
  \centering
  \makebox[\linewidth][c]{
  \includegraphics[trim=0 0 0 0,clip,width=1.1\linewidth]{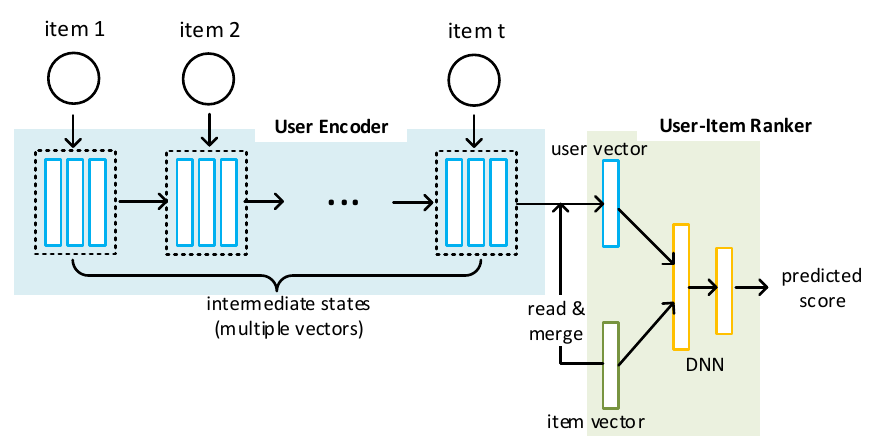}
  }
  \caption{A sequential user modeling framework with multiple vectors in intermediate states.}
\label{fig:model_pre}
\end{figure} 

\section{Problem Formulation}
Let $\mathcal{U}  = \{u_1, u_2, ..., u_n\} $ denote the set of users and $\mathcal{V}  = \{v_1, v_2, ..., v_m\}$ denote the set of target items, where $n$ and $m$ indicate the number of users and items respectively. Each user is associated with a sequence of behaviors: $B(u)=\{(x^u_1, t^u_1), (x^u_2, t^u_2), ..., (x^u_{|B(u)|}, t^u_{|B(u)|})\}$, where $u \in \mathcal{U}$ indicates a user, $x \in \mathcal{X}$ indicates an activity, $(x^u_1, t^u_1)$ indicates a user $u$ has an activity $x^u_1$ at time $t^u_1$, and $|B(u)|$ denotes the number of elements in the set. Behaviors are sorted by timestamps, i.e., we have $t_i <= t_j$ for any $i<j$. Items in $\mathcal{V}$ are not necessary to be the same as items in $\mathcal{X}$. For example, in our ads display scenario, $\mathcal{X}$ consists of web pages that users visited previously and $\mathcal{V}$ consists of advertisements to display. Each item $v$ and user behavior $x$ will be encoded into a $D$-dimensional vector: $\mathbf{v}, \mathbf{x} \in \mathbb{R}^{D}$.  

The recommendation task can be formulated as to predict a user $u$'s preference to item $v$ at time $t$: $\hat{y}=f(y |u, v, t)$, given the user's behavior history before time $t$. The key component lies in user modeling, i.e., how to generate a user representation $\mathbf{u}_t$ based on behavior sequence. As stated in Section \ref{sec:nrt}, to make the model scalable for online serving, we only consider the architectures which can be updated \textbf{incrementally}, such as GRU \cite{hidasi2015session} and RUM \cite{chen2018sequential}, which we call \textsl{User Encoder}. At any time $t$, we can readout a user vector $\mathbf{u}_t$ from the \textsl{User Encoder}. $\mathbf{u}_t$ will be concatenated with the candidate item vector $\mathbf{v}$ (and other context features if we have), then goes through a two-layer fully connected neural network (FCN) to get the prediction score:  $f(y |u, v, t) = FCN^2(\mathbf{u}_t, \mathbf{v})$. We call this part as \textsl{User-Item Ranker}. The model architecture is illustrated in Figure \ref{fig:model_pre}, where the two components can be matched accordingly in Figure \ref{fig:system} Step 3.

\section{The Proposed Method}
\label{sec:ourmodel} 
\begin{figure}[t]
\includegraphics[trim=200 130 250 140,clip,width=1.2\linewidth]{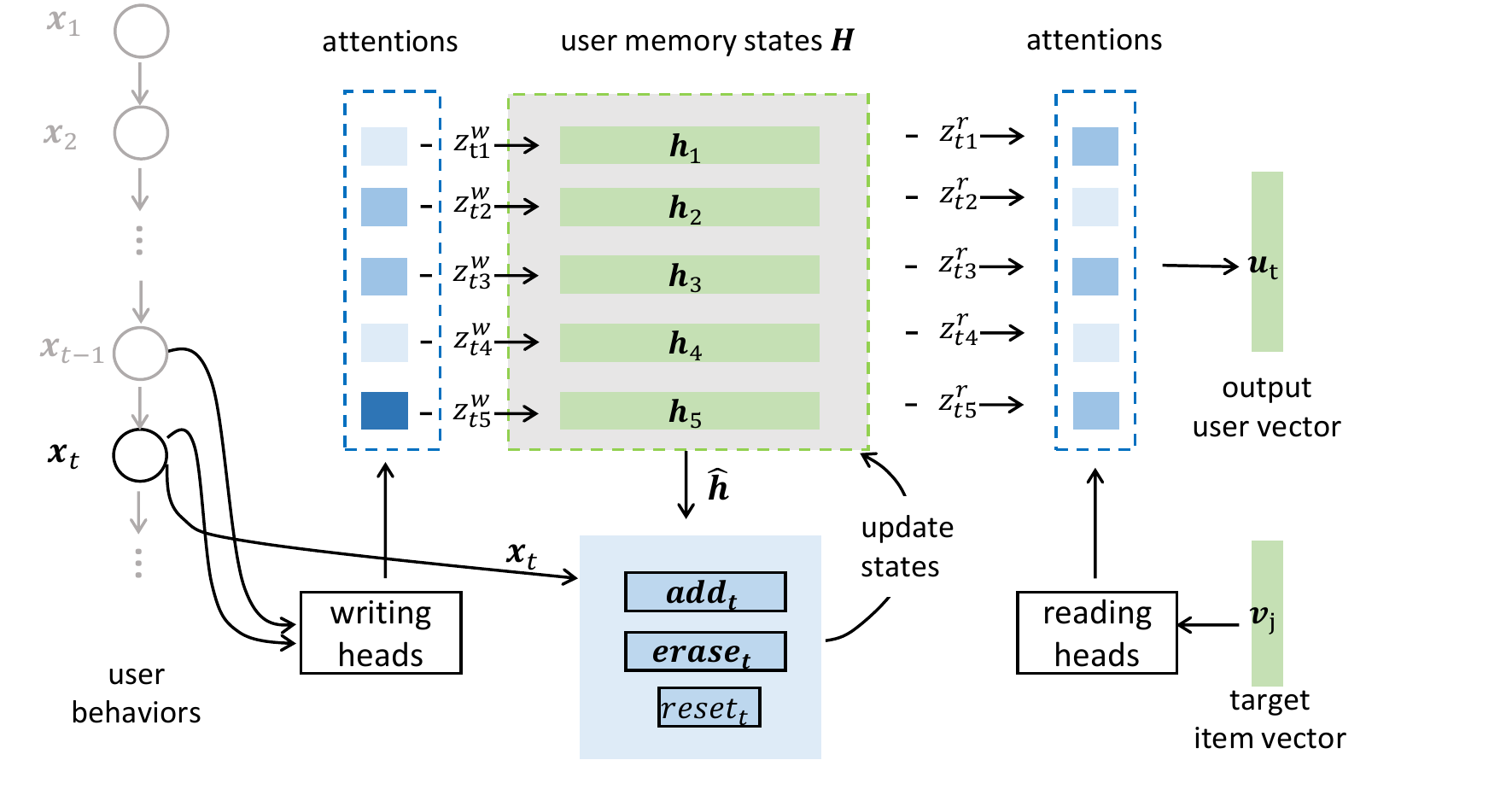}
\vspace{-0.4in}
\caption{An illustration of the architecture of SUM with 5 memory channels.}
\label{fig:model_sum}
\end{figure}

To effectively capture users' multiple interests in a scalable way for online serving, we propose Sequential User Matrix (SUM), which is built on memory networks \cite{chen2018sequential}, especially inspired by the Recommender model with User Memory network~(RUM)\cite{DBLP:journals/corr/GravesWD14}. We use memory states with K channels\footnote{In this paper, ``channels'' and ``slots'' are used interchangeably.} to represent a user: $\mathbf{H}^u = \{\mathbf{h}_1, \mathbf{h}_2, ..., \mathbf{h}_K\} \in \mathbb{R}^{D \times K}$. Figure \ref{fig:model_sum} illustrates how SUM processes a user's behavior sequence. There are basically two groups of operations: the writing operation and the reading operation. The writing operation takes one user behavior at a time as input and updates the memory states accordingly. The reading operation merges the memory states into one user vector for user-item preference prediction. We propose three key mechanisms in SUM's writing operation, i.e., the interest-level and instance-level attention, the local proximity debuff, and the highway channel.
\subsection{Interest-level and Instance-level Attention}
\label{sec:attention}
We assume that users historical behavior items may be different from the target items. For example, in Display Ads dataset, users historical behaviors are their visited webpages, while the target items are ads. Thus, we have two sets of global feature maps, $\mathbf{F}^w = \{\mathbf{f}^w_1, \mathbf{f}^w_2, ... ,  \mathbf{f}^w_K$ \} and $\mathbf{F}^r = \{\mathbf{f}^r_1, \mathbf{f}^r_2, ... ,  \mathbf{f}^r_K\}$, for writing and reading operation respectively (also known as writing heads and reading heads). When a new user behavior $\mathbf{x}_t$ comes in, we first compute its attention weight to each channel:
\begin{equation}
\label{eq:w_tk}
    w^w_{tk} = \mathbf{x}_t \cdot \mathbf{f}^w_k, \ \ \   z^w_{tk} = \frac{exp(\beta w^w_{tk})}{\sum_j exp(\beta w^w_{tj})}, \ \ \ \forall k=1,2,..., K
\end{equation}
where $\beta$ is a scaling factor, $z_{tk}$ represents the attention score of event $ \mathbf{x}_t$ towards channel $k$. We call this attention \textsl{interest-level attention}. This step is the same with RUM\cite{chen2018sequential}. However, we observe that in RUM, the updating vectors, such as $\mathbf{add}_t$ and $\mathbf{erase}_t$, only depend on the input event $\mathbf{x}_t$. We argue that the updating vectors for memory channels should consider both the current input $\mathbf{x}_t$ and the current state $\mathbf{H}$ (we call it \textsl{instance-level attention}). Thus, we first merge the memory states by the writing attention scores $\mathbf{z}^w_{t}$:
\begin{equation}
\label{eq:h_hat}
    \mathbf{\hat{h}} = \sum_k z^w_{tk} \cdot \mathbf{h}_{k}
\end{equation}
The new value which will be added to the memory states is dependent on both the input $\mathbf{x}_t$ and the current states $\mathbf{H}$:
\begin{equation}
\label{eq:add02}
    \mathbf{add}_t  =  \phi (\mathbf{W}_a [ \mathbf{x}_t ,  reset \cdot \mathbf{\hat{h}}] + \mathbf{b}_a)
\end{equation}
Meanwhile, we have the reset gate and erase gate:
\begin{equation}
\label{eq:erase02}
    \mathbf{erase}_t = \sigma (\mathbf{W}_e   [ \mathbf{x}_t ,   \mathbf{\hat{h}}] + \mathbf{b}_e)
\end{equation}
\begin{equation}
\label{eq:reset_t}
    reset_t = \sigma (W_r   [ \mathbf{x}_t ,   \mathbf{\hat{h}}] + b_r)
\end{equation}
Note that  $ reset_t$ is a scalar to control how much the current states are involved to generate the new add-on value. $\mathbf{erase}_t, \mathbf{add}_t \in \mathbb{R}^{D}$. The memory states are updated by:
\begin{equation}
    \mathbf{h}_k \gets \mathbf{add}_t  \cdot  \mathbf{erase}_t \cdot z^w_{tk} +  \mathbf{h}_k  \cdot  (\mathbf{1} - \mathbf{erase}_t \cdot z^w_{tk})
    \label{eq:sum_update_h}
\end{equation} 
which means we first erase a portion of information from the current states, then add new values to them. $\mathbf{erase}_t$ is a weighting vector which controls the erasing level in a bit-wise level. $z^w_{tk}$ is an interest-level attention score which controls the degree of information change on channel $k$ at timestamp $t$. Essentially, Equation \ref{eq:sum_update_h} is a linear interpolation of previous states $\mathbf{h}_k$ and the new add-on vector $\mathbf{add}_t $, we have the coefficient with $ \mathbf{erase}_t \cdot z^w_{tk} + (\mathbf{1} - \mathbf{erase}_t \cdot z^w_{tk}) = 1$ and $ 0 \leq \mathbf{erase}_t \cdot z^w_{tk} \leq 1 $. Since the current state $\mathbf{H}$ is reduced from previous user behaviors and impacted by writing heads, we call the state updating mechanism is of both \textsl{instance-level} (Eq.(\ref{eq:add02}, \ref{eq:erase02}, \ref{eq:reset_t})) and \textsl{interest-level} (Eq.(\ref{eq:w_tk})) awareness.
\subsection{Local Proximity Debuff}
\label{sec:lpd}
Users' consecutive behaviors tend to be similar. For example, when people search for some information (Nike shoes), a few pages/items she clicks on in a session usually belong to the same topic (Nike shoes). We call this phenomenon \textsl{local proximity}. This phenomenon is especially obvious when we are handling the original user behavior dataset without down-sampling or de-duplicating. Figure \ref{fig:case_study} in Section \ref{sec:casestudy} also verifies this. In online serving scenario, since the model is expected be updated incrementally in a streaming manner, it is not practical to store the user sequences and perform the de-duplication process. To alleviate the local proximity problem and make the model capable to remember long term user interest, we propose a simple but effective debuff \footnote{The terminology \textsl{debuff} originates from gaming. It means an effect that makes a game character weaker. \url{https://en.wiktionary.org/wiki/debuff}. We borrow this word to describe that we are weakening the local proximity effect.} mechanism based on the comparison between the current event $\mathbf{x}_{t}$ and the last event $\mathbf{x}_{t-1}$:
\begin{equation}
    sim^{(t)} = cosine(\mathbf{x}_t, \mathbf{x}_{t-1})
\end{equation}
\begin{equation}
    z^w_{tk} \gets z^w_{tk} \cdot \alpha ^{sim^{(t)} }
\end{equation}
Where $\alpha$ is a trainable parameter; we initialize it with a value slightly smaller than 1.0, such as 0.98. When $\alpha$ is less than 1, the more similar a pair of consecutive events ($\mathbf{x}_t, \mathbf{x}_{t-1}$) are, the smaller the attention score $z^w_{tk}$ will be. Although we have different choices for the debuff mechanism, such as feature based approach like $sim^{(t)} = FCN(\mathbf{x}_t, \mathbf{x}_{t-1}, \mathbf{x}_t \odot \mathbf{x}_{t-1}, \mathbf{x}_{t} - \mathbf{x}_{t-1})$, our method doesn't bring in additional parameters except a single scalar $\alpha$, which ensures a more fair comparison between methods with/without the local proximity debuff mechanism. Through experiments (Section \ref{sec:ablation}) we find that this method works very well.

\subsection{Highway Channel}
\label{sec:hw}
Although memory networks have the stronger expressive capacity in general than recurrent neural networks (RNN) such as GRU, there is one thing that RNN is theoretically better than memory networks: RNN has the ability to record the order of events in sequence explicitly. In contrast, in memory networks, the original order of events is not strictly maintained because the coming event is routed to different channels with different attention weights. The events' order is critical for sequential recommendations since the most recent event usually plays the most important role for the next item prediction. Motivated by this, we reserve one memory channel to be a \textsl{highway channel}, which always has an attention weight of 1 for all the coming events. Since all the events have interacted on this channel evenly, we empower the memory network to capture the exact order of events. Including the highway channel can make the SUM model more robust over various datasets. 
The degree of user interest diversity varies over different datasets. The highway channel represents a mixture of interests, while the other memory channels represent a disentanglement of interests.  
The union of the two types of channels empowers the SUM model with the flexibility of switching between interest mixture and interest disentanglement adaptively with different datasets.   

\subsection{Reading Operation}
\label{sec:readingop}
To read the user memory matrix $\mathbf{H}$ at time $t$, \cite{chen2018sequential} first uses a global reading latent feature table $\mathbf{F}^r = \{\mathbf{f}^r_1, \mathbf{f}^r_2,  ... , \mathbf{f}^r_K$ \} to get the candidate item $j$'s attention weight with each channel, then generates a user vector $\mathbf{u}_t$ by weighted average.  The equations are as follows:
\begin{equation}
\label{eq:old_read_att}
    w^r_{tk} = \mathbf{v}_j \cdot \mathbf{f}^r_k
\end{equation}
\begin{equation}
    z^r_{tk} = \frac{exp(\beta w^r_{tk})}{\sum_d^K exp(\beta w^r_{td})}, \ \ \ \forall k=1,2,..., K
\end{equation} 
\begin{equation}
\label{eq:u_t}
    \mathbf{u}_t = \sum_k {z}^r_{tk} \cdot \mathbf{h}_{k}
\end{equation} 
However, through experiments we found that this reading operation is not very effective. We argue that the attention weights $w^r_{tk}$ should depend on the content of user memory states. If a memory channel is rarely activated in the past, it should be assigned with a low attention score in the reading operation. We still use an attentive merging approach with Eq.(\ref{eq:u_t}) to read the user states, but change Eq.(\ref{eq:old_read_att}) to:
\begin{equation}
    w^r_{tk} = \mathbf{v}_j \mathbf{F}^r \mathbf{h}_{k}
\end{equation}
where $\mathbf{F}^r \in \mathbb{R}^{D\times D}$ is a global reading transformation matrix. We will report the comparison results in Section \ref{sec:ablation}.
 
\subsection{Learning}
After we got the representation $\mathbf{u}_{t, i}$ for user $i$, we concatenate it with the target item $j$'s representation $\mathbf{v}_j$, feed it into  a 2-layer fully-connected neural network with ReLU activation function, then connect it to an output preference score unit which Sigmoid activation function.
Since the focus of this paper is about sequential user modeling, we do not include more features (such as context features) or use more complicated scorer (such as \cite{DBLP:conf/ijcai/GuoTYLH17}) in this prediction process for the sake of simplicity. However, richer features and modeling architectures can be easily included in this step.
We take the user preference prediction problem as a binary classification task, so in this paper, we use a point-wise loss with a negative log-likelihood function:
\begin{equation}
    \mathcal{L} =-\frac{1}{N}\sum_{i, j}y_{i,j} log \hat y_{i,j} + (1-y_{i,j})log(1-\hat y_{i,j}) + \lambda_||\Theta||^2
\end{equation}
where $N$ is the total number of training instances, $\Theta$ denotes the set of trainable parameters.

\subsection{Complexity Discussion}
As for the inference computational cost, our SUM model is on the same level with GRU. The most expensive step of GRU is calculating the new state in forms of $\sigma(\mathbf{W} [\mathbf{x}_t, \mathbf{h}_{t-1}] + \mathbf{b}_z)$, the time complexity is $O(D^2)$. Although in SUM there are K channels, from Eq (\ref{eq:add02}) - Eq (\ref{eq:reset_t}) we can see that updating vectors are shared among K channels. Only Eq (\ref{eq:h_hat}),(\ref{eq:w_tk}),(\ref{eq:sum_update_h}) are repeated for K channels, but their time complexity is $O(D)$ which is much less than $O(D^2)$.

\section{experiments}
\label{sec:exp}

\subsection{Datasets}
We use two datasets for experiments, including a display ads dataset and an e-commerce item recommendation dataset. Some basic data statistics are reported in Table \ref{tab:dataset_stat}. 

\textbf{Display Ads Dataset}. We collect two weeks' ad clicking logs from the Bing Native Advertising service as data samples and collect users' web behavior history before their corresponding ad click behavior for user modeling. The user behavior sequences are truncated to 100. The data samples are split into 70\%/15\%/15\% as training/validation/test dataset by users to avoid information leakage caused by repeated user behaviors. Both items and user behaviors are described by textual content. We use CDSSM \cite{shen2014latent} model as a text-encoder to turn the raw text into a 128-dimension embedding vector. The embedding vector is then used as the static feature for a user page view behavior.

\textbf{Taobao Dataset}. This is a public e-commerce dataset\footnote{\url{https://tianchi.aliyun.com/dataset/dataDetail?dataId=649&userId=1}} collected from Taobao's recommender system. To make two experimental datasets coherent, we take the purchase behaviors as target activities (which corresponds to the ads clicking behavior in Display Ads Dataset) and use page view behaviors for user modeling data (which corresponds to the web browsing behavior in Display Ads Dataset). The user behavior sequences are truncated to 100.  Since we don't have the non-click impression logs in this dataset, all the negative instances are randomly sampled according to item popularity with a positive:negative ratio of 1:4.  So we use item id and category id as one-hot features to represent items. 

For more detailed descriptions related to datasets, please refer to Appendix \ref{sec:dataset_appendix}.

\begin{table}
	\centering
	\caption{Statistics of the datasets. ``k'' indicates a thousand.}
	\vspace{-0.15in}
    \label{tab:dataset_stat}
	\begin{tabular}{c|ccccc}
		\toprule
		 Dataset & \#.Users & \#.Items & \# \makecell{Positive \\ Instances}  & \makecell{Avg Length of  \\  user behaviors}   \\
		\midrule
		Display Ads & 748k & 409k & 1,024k   & 74    \\ 
		Taobao      & 623k & 554k &   1,868k  & 67   \\ 
		\bottomrule
	\end{tabular}
	\vspace{-0.15in}
\end{table}

\subsection{Baselines}
We compare SUM with three groups of methods:
\begin{itemize}[leftmargin=*]
    \item \textbf{AttMerge}. It represents users by attentively merging the historical items. It is non-sequential.
    \item \textbf{GRU} \cite{hidasi2015session,hidasi2018recurrent,10.1145/3097983.3098108}, \textbf{SGRU} and \textbf{HRNN} \cite{10.1145/3109859.3109896} represent GRU-based baselines.  
     SGRU stands for the \textsl{Stacked GRU}, which has multiple layers of GRU, with the layer number equivalent to the slots of SUM. Since GRU is a single vector-based method, SGRU is a more fair baseline to compare with SUM. \textsl{HRNN} is a hierarchical GRU with sequences organized by sessions.
     
    \item \textbf{NTM} \cite{DBLP:journals/corr/GravesWD14}, \textbf{RUM} \cite{chen2018sequential},  \textbf{MCPRN} \cite{wang2019modeling},  \textbf{MIMN} \cite{pi2019practice}, \textbf{HPMN} \cite{ren2019lifelong} are various kinds of multi-channel-based sequential user models, which represent a set of strong baselines. For all these models as well as SUM, we fix the channel number to 5 for fair comparisons.
\end{itemize}
All models share the same \textsl{User-Item Ranker} module marked with the green part in Figure \ref{fig:model_pre}, while each model uses its own blue \textsl{User Encoder} module. User states sizes are 128 for Ads dataset and 64 for Taobao dataset.
In this paper we only study models which can be updated incrementally. Thus, some other popular methods, such as DIEN \cite{zhou2019deep}, SASRec \cite{DBLP:conf/icdm/KangM18}, MIND \cite{li2019multi} and ComiRec \cite{cen2020controllable} are not listed as baselines.  
Hyper-parameter settings are listed in Appendix \ref{sec:parameters_appendix}.

\begin{table}[t]
\renewcommand{\tabcolsep}{1mm}
	\centering
	\caption{Overall performance comparison in terms of Group AUC, LogLoss and NDCG@3. A \textbf{bold} font means the number is significantly bigger than the second best model with $p$-value $<$ 0.05. For notation simplicity we omit the asterisks.}
	\label{tab:overall_res}
	\vspace{-0.15in}
	\begin{tabular}{c|ccc|ccc} 
	\hline
	    & \multicolumn{3}{c|}{Display Ads}    & \multicolumn{3}{c}{Taobao}       \\
		Model & gAUC & LogLoss & NDCG & gAUC & LogLoss & NDCG \\
		\midrule
		AttMerge   & 0.7788  & 0.4144  &   0.6784          &  0.8978 &  0.2752  & 0.8655    \\ 
		\midrule  
		GRU  &  0.8262 & 0.3768  &     0.7306        & 0.9279  & 0.2133   & 0.9052  \\
		SGRU  &  0.8250 & 0.3781  &  0.7296           &  0.9360 & 0.2054   & 0.9152  \\
		HRNN  & 0.8284  & 0.3769  &  0.7334           & 0.9267  &  0.2168  & 0.9036  \\
		\midrule  
		MCPRN  & 0.8258  & 0.3789  &  0.7311           &  0.9348 &  0.2066  &  0.9144  \\
		NTM  &  0.8256 & 0.3786  &  0.7302           & 0.9251  &  0.2190  & 0.9014  \\
		RUM  &  \underline{0.8303} & \underline{0.3742}  &   \underline{0.7366}          &  0.9300 & 0.2197 & 0.9011 \\
		MIMN  &  0.8280 &  0.3755 & 0.7327            & 0.9286  &  0.2137  &  0.9060 \\
		HPMN  & 0.8262  & 0.3769  &   0.7313          & \underline{0.9361}  &  \underline{0.2033}  &  \underline{0.9161} \\
		\midrule
		SUM & \textbf{0.8342} & \textbf{0.3719} & \textbf{0.7409 } & \textbf{0.9420} & \textbf{0.1896}  &  \textbf{0.9235}  \\ 
	\hline
	\end{tabular} 
\end{table}

\subsection{Evaluation Metrics}
We adopt three widely-used metrics for evaluation: \textbf{Group AUC} (Area Under the ROC curve), \textbf{Logloss} (binary cross entropy) and \textbf{NDCG} (Normalized Discounted Cumulative Gain). From the user recommendations perspective, we only need to compare among the candidates for a given user. So we adopt the Group AUC, which first calculates a AUC score per user, then takes the average among users.  Logloss measures the distance between the predicted score and the true label for each instance, which is also frequently used in recommendation task \cite{DBLP:conf/ijcai/GuoTYLH17,10.1145/3219819.3220023,ren2019lifelong}. NDCG measures the ranking quality among top-k predicted candidates. We observe the same trend for different k among compared models, for conciseness, we only report NDCG@3 in the experiment section.

\subsection{Overall Performance Comparison} 
We first compare the overall performance of SUM with the aforementioned competitive methods. The results are reported in Table \ref{tab:overall_res}. We make the following observations.

Among all the benchmark methods, \textsl{AttMerge} is the only one that is non-sequential. As we focus on the sequential recommendation scenario, the fact that \textsl{AttMerge} falls far behind the other methods is expected.

\textsl{GRU} is a strong baseline method and is so far most widely used in industry as incrementally updatable. \textsl{SGRU} and \textsl{HRNN} are two updated versions of GRU-based models, which have more complicated structures than \textsl{GRU}. However, both \textsl{SGRU} and \textsl{HRNN} fail to consistently outperform \textsl{GRU} over two datasets, which indicates that simply stacking GRU layers or breaking behavior sequences into sessions are not powerful and generalized enough to handle various kinds of datasets. 
    Moreover, If we directly apply the vanilla memory network architecture, i.e., \textsl{NTM}, for user modeling, the performance is worse than \textsl{GRU} models.  

\textsl{MCPRN}, \textsl{RUM}, \textsl{MIMN} and \textsl{HPMN} all leverage and improve memory networks for user modeling. Although the best one among them is better than GRU-based models when considering different datasets separately, none of them can beat all GRU-based models on both two datasets. For example, \textsl{HPMN} performs very well on the Taobao dataset, but it performs almost the same with \textsl{GRU} on the Display Ads dataset.

\textsl{SUM} outperforms all the baseline methods in different evaluation metrics significantly, which verifies the effectiveness of the proposed model. More importantly, \textsl{SUM} can perform best consistently on both datasets, which demonstrates the robustness of our proposed method.

\begin{table}[t]
	\centering
	\caption{Disabling every component in SUM will lead to a performance drop. }
	\label{tab:ablation_study}
	\vspace{-0.15in}
	\begin{tabular}{c|cc|cc}
	\hline
	    & \multicolumn{2}{c|}{Display Ads}    & \multicolumn{2}{c}{Taobao}       \\
		Model & gAUC & NDCG &   gAUC & NDCG \\
		\midrule
		SUM &   \textbf{0.8342 } & \textbf{0.7409 }   & \textbf{0.9420}  &  \textbf{0.9235}  \\ 
		\midrule  
		w/o instance-level att   &  0.8321 &   0.7389      &  0.9343  &  0.9138  \\
		w/o proximity debuff    & 0.8320   &    0.7385     &  0.9410  & 0.9222 \\
		w/o highway channel   &  0.8316  &  0.7378       &  0.939  &  0.9196  \\ 
		reading operation (-)   &  0.8320  &  0.7374      & 0.9348   & 0.9140  \\
		writing like NTM    &  0.8306  &   0.7354      & 0.9309   & 0.9090  \\
	\hline
	\end{tabular} 
\end{table}

\begin{table}[th]
	\centering
	\caption{LPD benefit comparison for GRU, RUM and SUM. }
	\label{tab:lpb_benefit}
	\vspace{-0.15in}
	\begin{tabular}{c|cc|cc}
	\hline
	    & \multicolumn{2}{c|}{Display Ads}    & \multicolumn{2}{c}{Taobao}       \\
		Model & gAUC & NDCG &   gAUC & NDCG \\
		\midrule
		GRU    &  0.8262 &  0.7306    &   0.9279  &  0.9052 \\
		GRU w/  LPD &  0.8267  & 0.7312   & 0.9273  &  0.9051  \\ 
			\midrule
		RUM    &  0.8303 &  0.7366    &  0.9300  &   0.9011 \\
		RUM w/  LPD &   \textbf{0.8324} & \textbf{0.7387}   & \textbf{0.9324}  &  \textbf{0.9113}  \\ 
			\midrule
		
		SUM w/o LPD   &  0.8320 &  0.7385     &  0.9410  &   0.9222 \\
		SUM &   \textbf{0.8342} & \textbf{0.7409}   & \textbf{0.9420}  &  \textbf{0.9235}  \\  
	\hline
	\end{tabular} 
\end{table}

\subsection{Ablation Study}
\label{sec:ablation}
Key components in SUM include the instance-level attention, the local proximity debuff (LPD), the highway channel, and the reading/writing attention mechanism. To verify each component's impact, we disable one component each time while keeping the other settings unchanged, then test how the performance will be affected. We use \textsl{reading operation(-)} to denote the alternative setting of the reading operation as stated in Section \ref{sec:readingop}. We further replace SUM's writing mechanism with NTM's writing mechanism, which is denoted as \textsl{writing like NTM}.  From Table \ref{tab:ablation_study} we can see that removing either one component from SUM will cause a consistent performance drop on both datasets.

Among the key components of SUM, perhaps the most incomprehensible one is the LPD. Thus, we conduct additional experiments to understand better the impact of LPD. LPD is a flexible unit that can be plugged into different models. We choose three models \textendash GRU, RUM and SUM \textendash and report the results in Table \ref{tab:lpb_benefit}. Interestingly, LPD benefits RUM and SUM, but it doesn't make a significant difference to GRU. The reason is that GRU has a gated mechanism to control how much information to absorb and how much memory to forget, and the gated attention can be well determined by comparing the hidden state and the input event. However, when it comes to multi-channel hidden states, it is hard to sense the local proximity information due to the feature that behaviors are dispersed to different channels by user interest. Therefore, LPD is beneficial to multi-channel-ware models like RUM and SUM, but it is meaningless to a single-channel model like GRU.

\begin{figure}[t]
\begin{subfigure}{.5\linewidth}
  \centering
  \includegraphics[trim=0 0 0 0,clip,width=0.95\linewidth,height=.8\linewidth]{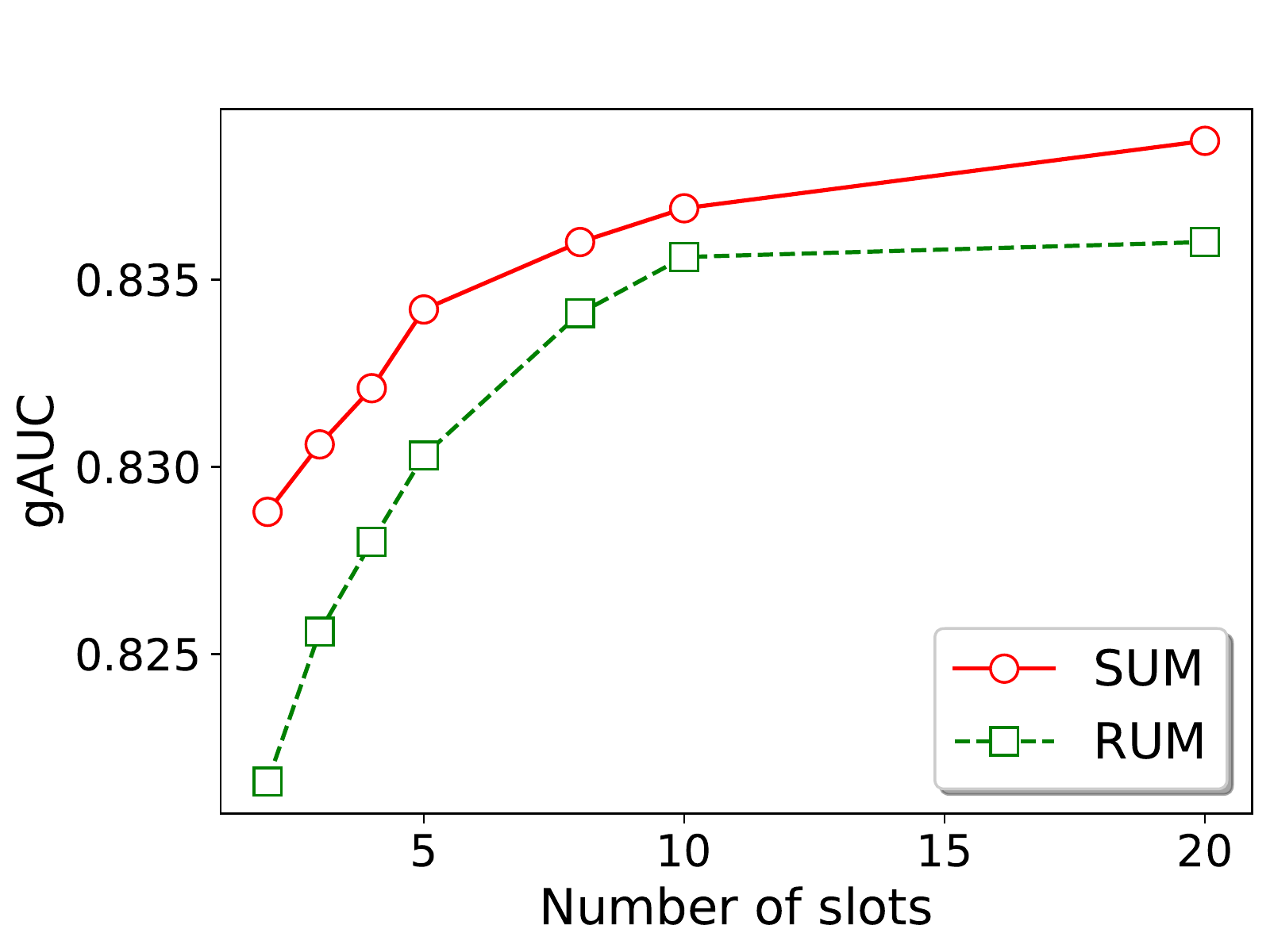}   
  \vspace{-0.1in}
  \caption{ Display Ads.} 
\end{subfigure} 
\hspace{-1em}
\begin{subfigure}{.5\linewidth}
  \centering
  \includegraphics[trim=0 0 0 0,clip,width=0.95\linewidth,height=.8\linewidth]{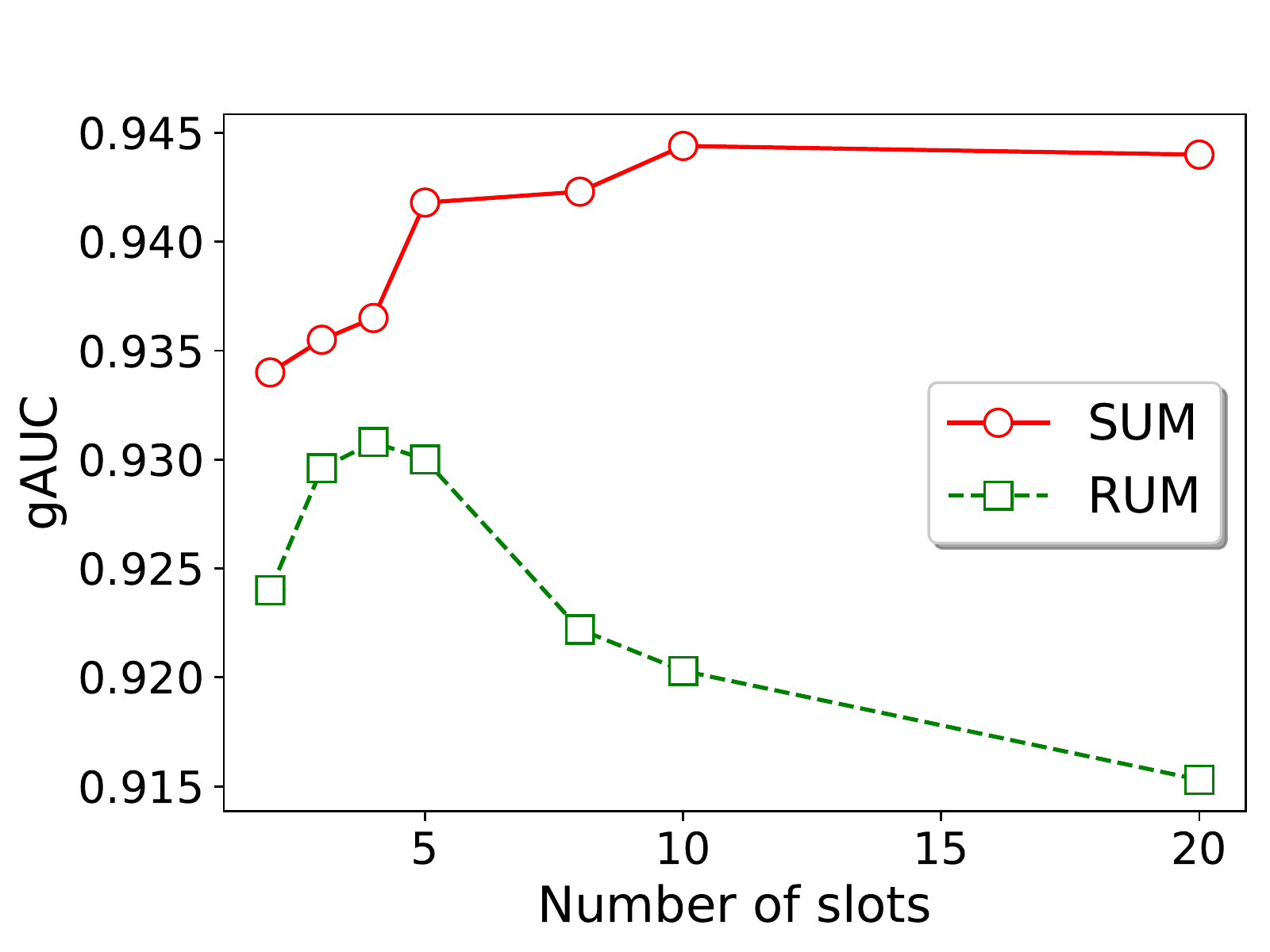}
  \vspace{-0.1in}
  \caption{Taobao.} 
\end{subfigure} 
\vspace{-0.15in}
\caption{Performance with different number of slots.}
\label{fig:impact_slots}
\end{figure}

\subsection{Impact of Number of Channels}
\label{sec:exp_channel}
Figure \ref{fig:impact_slots} demonstrates how the number of memory channels impacts SUM's performance. 
For comparison, we also plot the lines of RUM to Figure \ref{fig:impact_slots}.
We observe that performance patterns on the two datasets are slightly different. For the Display Ads dataset, a good setting for the channel number is 10, and further increasing the channel number does not improve the accuracy significantly. However, for the Taobao dataset, the ideal channel number is around 5, after which SUM' performance becomes saturated while RUM starts to decline. We learn that the model capacity can not be enhanced infinitely by adding more channels through this experiment. Allocating excessive channels will make the model hard to converge and maybe even damage the performance.

\begin{figure}[t]
  \centering
  \includegraphics[trim=20 0 0 20,clip,width=0.9\linewidth]{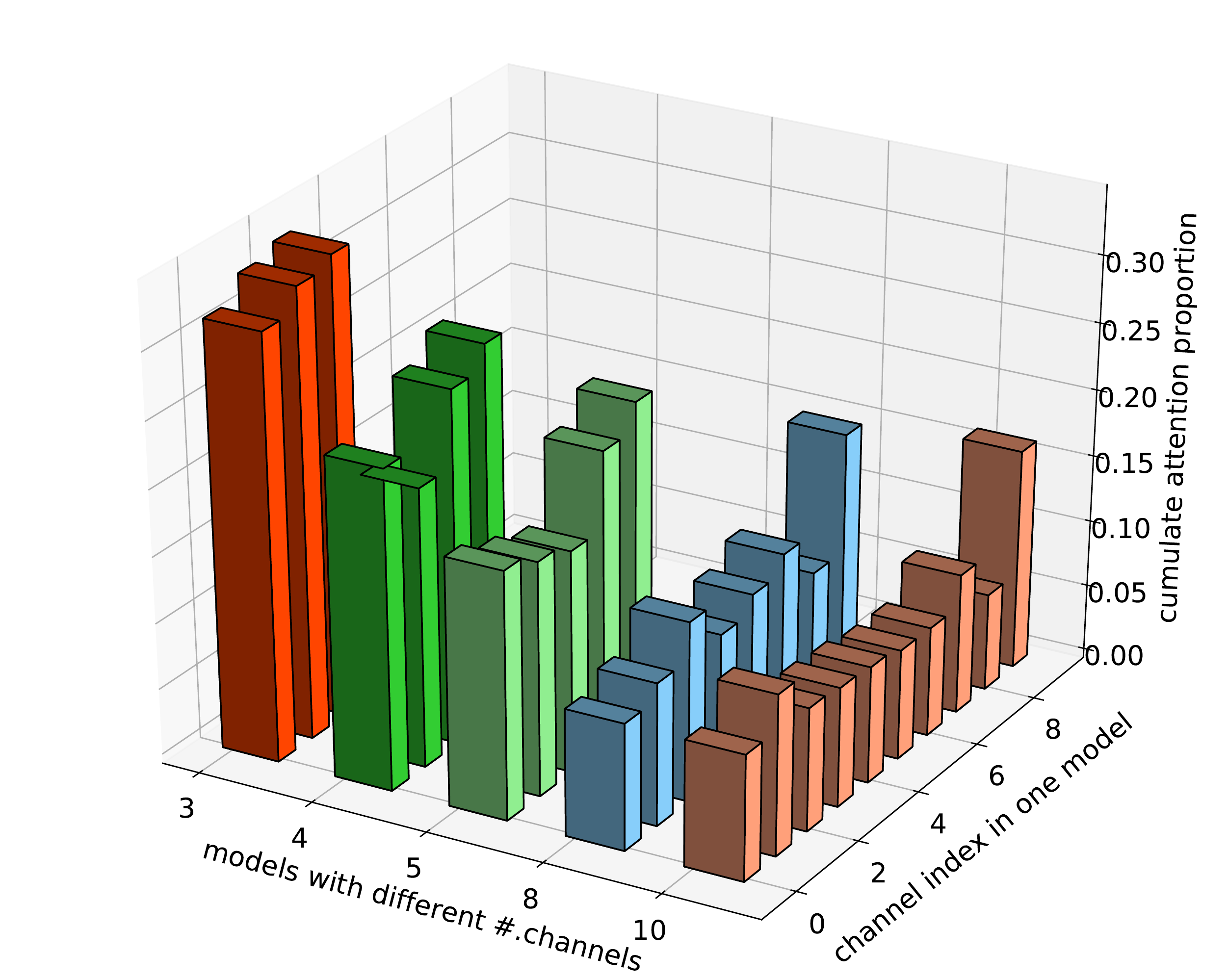}
  \vspace{-0.2in}
  \caption{The proportion of cumulative readout attentions over all users for each channel. We analyze 5 models with different setting of channel number in \{3, 4, 5, 8, 10\}. For each model, the last index of channel corresponds to the highway channel. Dataset is the display ads.}
\label{fig:cumulative_att}
\end{figure}

\subsection{Channel Utilization} 
Next we analyze the utilization of channels in one model. Note that we have two types of attentive operations in SUM, i.e., the writing attention and the reading attention. For the writing attention pattern study, we report the average number of activated channels per user in the writing stage. Here, ``an activated channel'' is defined as the channel with the largest writing attention score for user behavior. Take a SUM model with 5 channels as an example. After going through the behavior sequence of one user, if the activated channel number is 2, then the utilization  for her is $2/(5-1)=0.5$. Note that the highway channel is not counted. Table \ref{tab:channel_util} summarizes the patterns. Overall, the writing utilization is high, and as the number of channels increases, the utilization ratio drops. As for the reading attention pattern study, we plot Figure \ref{fig:cumulative_att}. We sum up the readout attention scores over all the positive instances for each channel, then normalize the cumulative attention scores across channels. We want to discover how every channel takes effect on average. Again we try different SUM setting with channel number in \{3, 4, 5, 8, 10\}. For every experiment, the last index of channels denotes the highway channel. If the channels are randomly utilized, every channel's cumulative attention proportion will be around $1/10=0.1$. An interesting finding from Figure \ref{fig:cumulative_att} is that, overall, the channels' reading utilization is even, and as the channel number increases, the highway channel's utilization proportion becomes more prominent. This phenomenon is reasonable. With the increment of channel number, each channel will store more fine-grained interest for users. The distinction between the highway channel and the other channels become clearer, and the highway channel will play a more important role in modeling users' integrated interest.

\begin{table}
	\centering
	\caption{Writing utilization on the Display Ads dataset.}
	\vspace{-0.15in}
    \label{tab:channel_util}
	\begin{tabular}{c|c|c|c|c|c} \hline
		 \#.Channels & 3   & 4  & 5 & 8 & 10 \\ \hline
		AVG Utilization & 0.975 & 0.946 & 0.925   & 0.845 & 0.803    \\    \hline
	\end{tabular}
	\vspace{-0.15in}
\end{table}

\begin{figure*}[t]
\begin{subfigure}{.95\textwidth}
  \centering
  \includegraphics[trim=0 160 0 180,clip,width=0.9\linewidth]{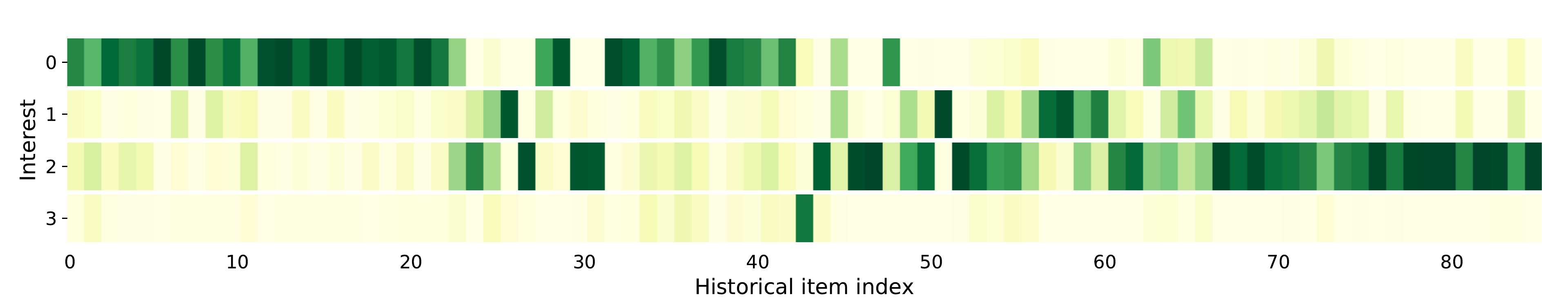}
  \vspace{-0.05in}
  \caption{ User A from the Taobao dataset. } 
\end{subfigure} 
\hspace{0.5em}
\begin{subfigure}{.95\textwidth}
  \centering
  \includegraphics[trim=0 140 0 180,clip,width=0.9\linewidth,,height=2.4cm]{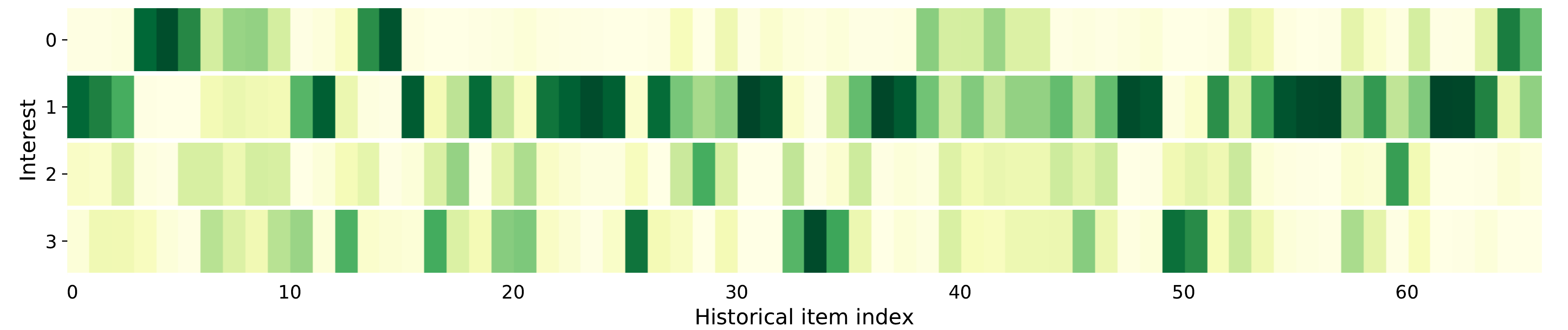}
    \vspace{-0.05in}
  \caption{User B from the Taobao dataset.} 
\end{subfigure} 
\vspace{-0.15in}
\caption{Heat map of channel coefficients of behavior sequences from two randomly sampled users in the Taobao dataset.}
\label{fig:case_study}
\end{figure*}

\begin{figure*}[t] 
\begin{subfigure}{.24\textwidth}
  \centering
  \includegraphics[width=0.99\linewidth]{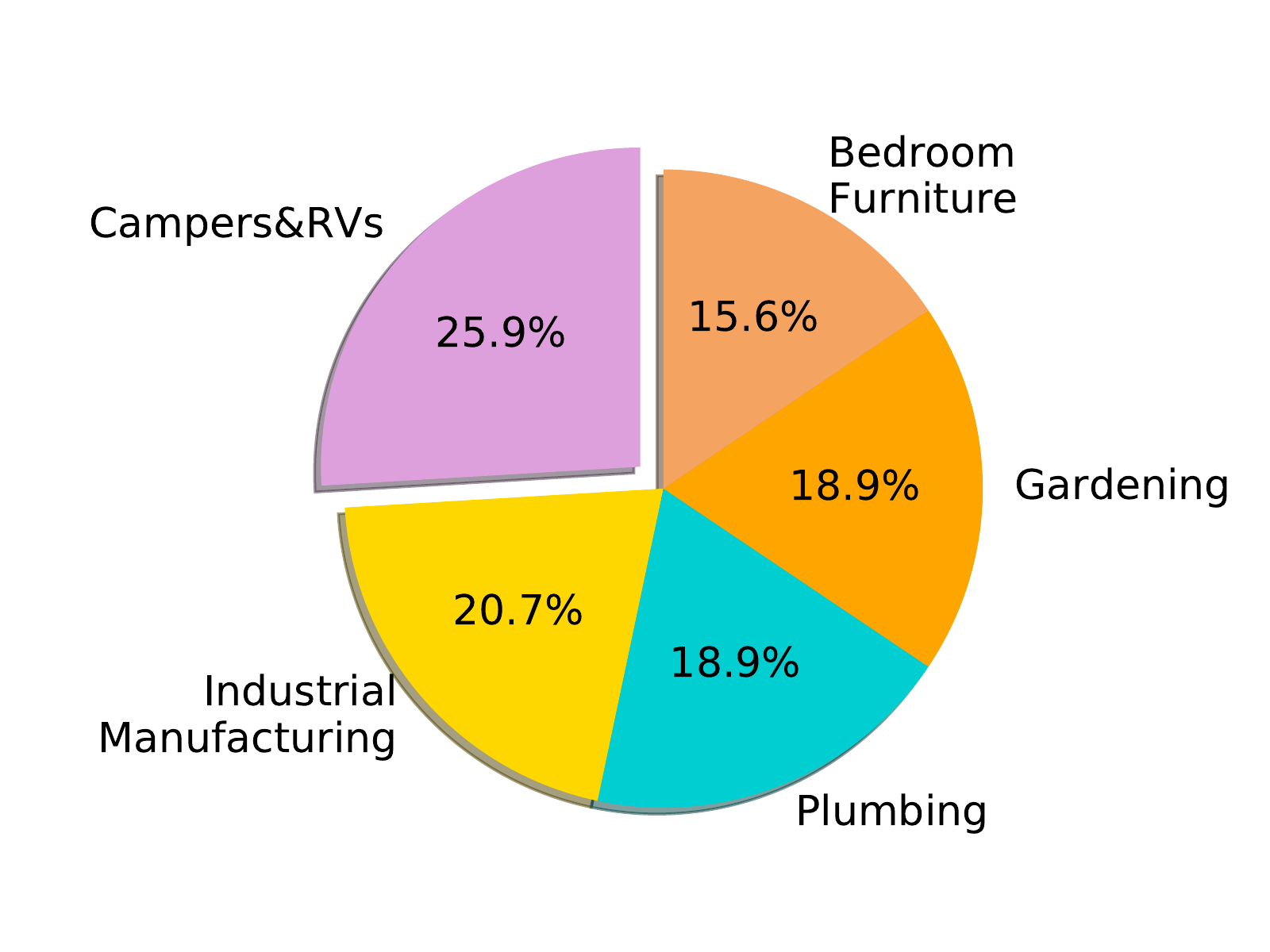}
    \vspace{-0.15in}
  \caption{ Channel 0. } 
\end{subfigure} 
\begin{subfigure}{.24\textwidth}
  \centering
  \includegraphics[width=0.99\linewidth]{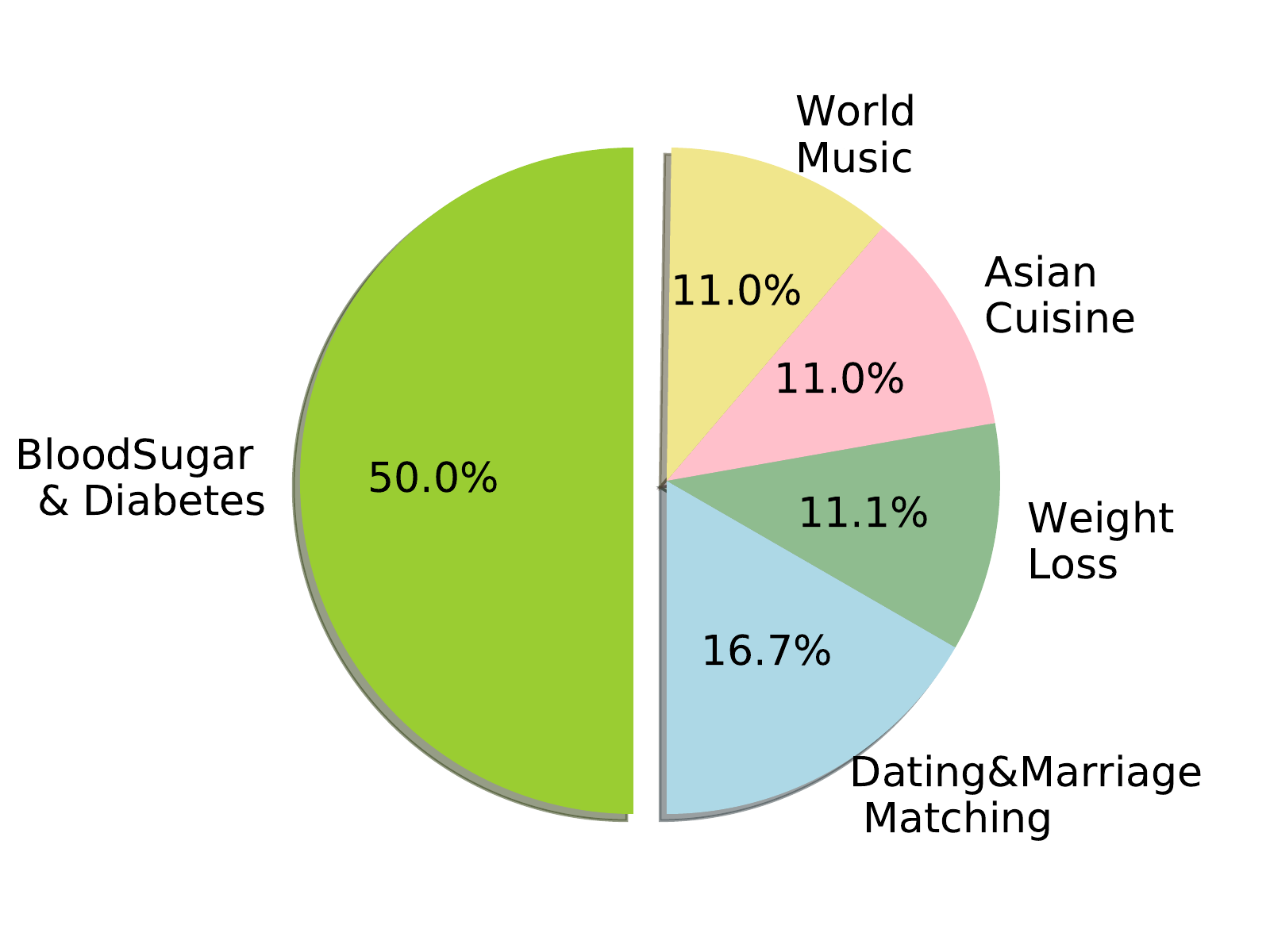}
    \vspace{-0.15in}
  \caption{  Channel 1.  } 
\end{subfigure} 
\begin{subfigure}{.24\textwidth}
  \centering
  \includegraphics[width=0.99\linewidth]{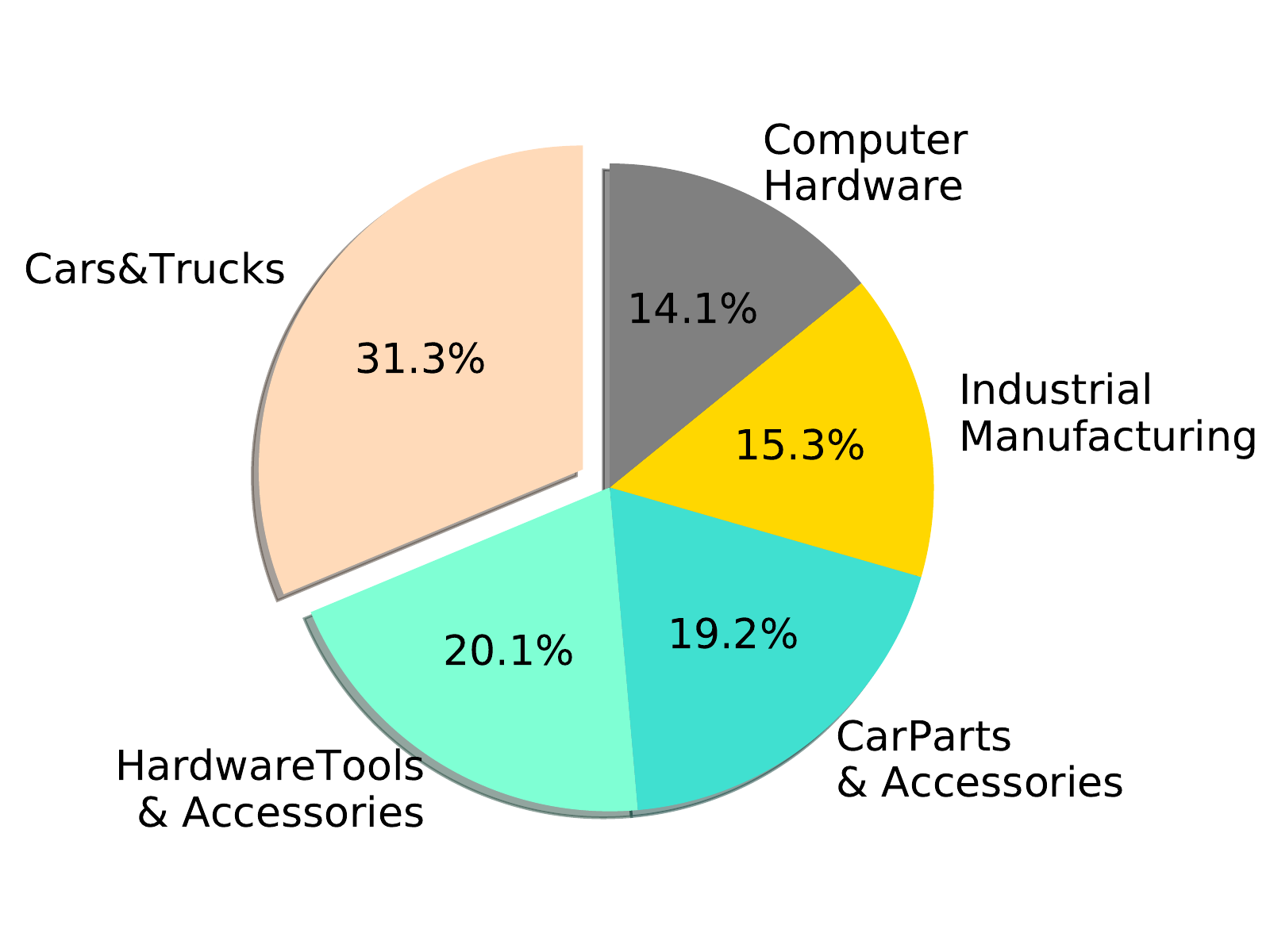}
    \vspace{-0.15in}
  \caption{  Channel 2.  } 
\end{subfigure} 
\begin{subfigure}{.24\textwidth}
  \centering
  \includegraphics[width=0.99\linewidth]{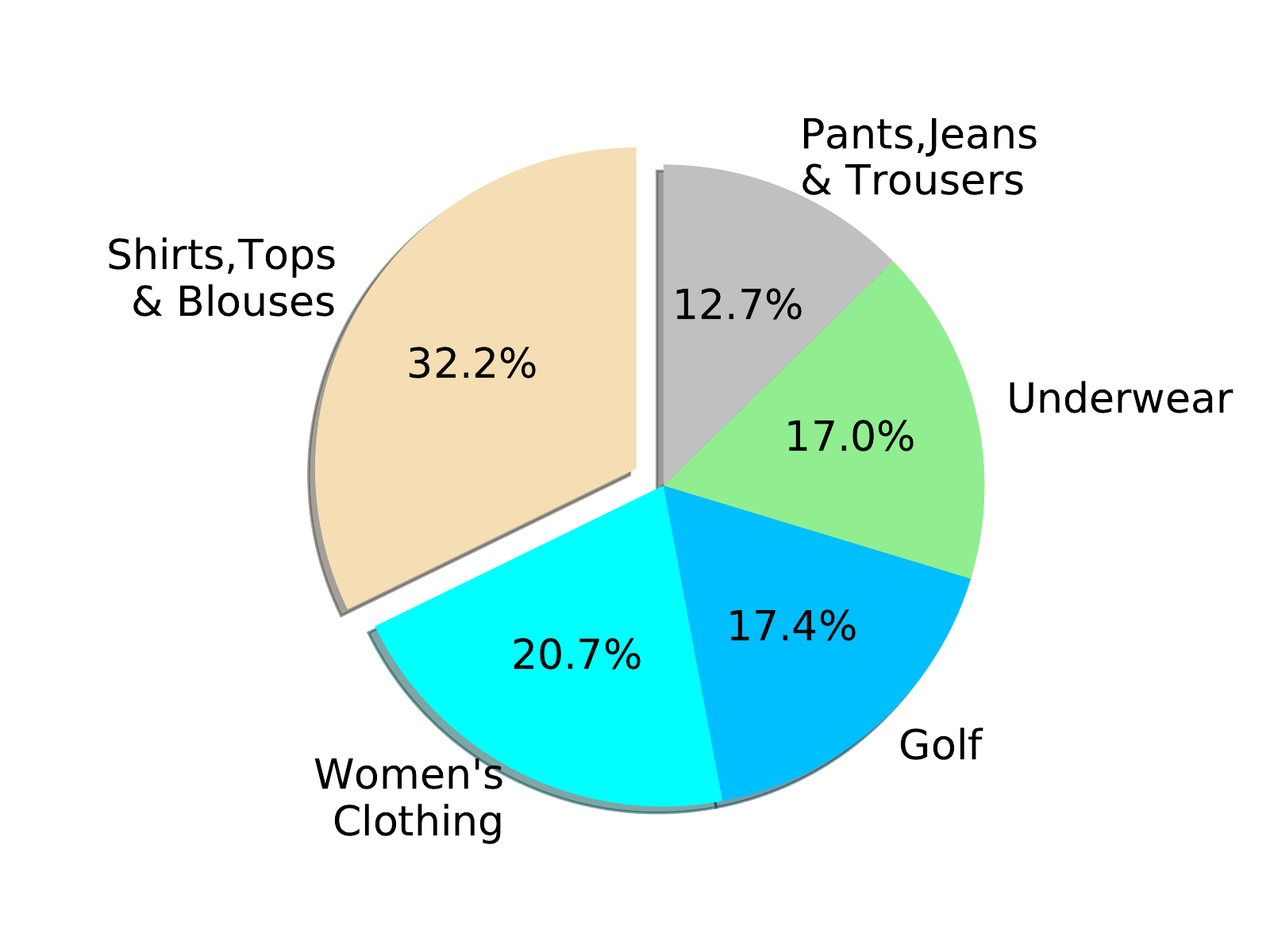}
    \vspace{-0.15in}
  \caption{  Channel 3.  } 
\end{subfigure} 
\vspace{-0.15in}
\caption{Percentages of top item categories for each channel on the Display Ads dataset. For each channel, the most frequent category is exploded for better illustration.}
\label{fig:case_study_channel_ads}
\end{figure*}

\subsection{Case Studies} 
\label{sec:casestudy}
We provide some case studies to demonstrate how user behaviors are distributed to different interest channels. We randomly sample two users from the Taobao dataset and report the results in Figure \ref{fig:case_study}. A darker color indicates higher attention scores on a channel. We report the attention scores on 4 channels, excluding the highway channel because it always has an attention score of 1. We can observe that (1) users have different behaviors on the four interest channels. User A's preference seems to evolve from interest 0 to interest 2, and her interests are mostly concentrated on channel 0 and channel 2. In contrast, user B's interests look more evenly distributed; And (2) user behaviors indeed have the local proximity property, as events with darker colors are prone to emerge in series.  To get a concrete sense of what different channels like, we plot Figure \ref{fig:case_study_channel_ads} with our Display Ads dataset (note that the Taobao dataset is anonymized so that we cannot get the real meaning of each category), on which we know the real meaning of each category and thus can clearly differentiate the latent interests of each channel. For example, channel\#0 covers a lot of items related to \textsl{Home \& Garden}, channel\#1 covers more things about \textsl{Health}, channel\#2 frequently mentions \textsl{Vehicles}, and top items in channel\#3 are related to \textsl{Apparel}. Interestingly, every channel is a mixture of various item categories, instead of being dedicated to one category. 

\subsection{Online Experiments}
We have deployed SUM in our NRT system for native ads serving. We conduct online A/B test on one of our main native ads traffic, with the treatment group being SUM and the control group being GRU, because GRU is our best prior production model. After a period of 23 days, the treatment group achieves 1.46\% gain in click yield (clicks over page views) and 1.32\% gain in revenue. We are still in the progress of making SUM generalized to all different traffic slides before it can fully replace our existing production model.

\section{Related works}
The sequential recommender system is an important branch of recommender systems that has attracted extensive attention in recent years.  \cite{hidasi2015session,hidasi2018recurrent,wu2017recurrent} are some good early works to discuss applying RNN on recommender systems motivated by successful application of RNN in other domains like natural language understanding.  \cite{tang2018personalized} proposes to use convolutional neural networks to model user behavior sequences and \cite{10.1145/3289600.3290975} further improves it by leveraging a dilated convolutional networks to increase the receptive fields. Motivated by the recent success of self-attention based models and pretrained models, some researchers proposed to leverage Transformer for sequential recommender systems \cite{10.1145/3357384.3357895,DBLP:journals/corr/abs-1808-06414,10.1145/3336191.3371786,DBLP:conf/icdm/KangM18}.
In industry cases, \cite{10.1145/3097983.3098108} uses an RNN with GRU cell to generate user representations with user browsing histories as input sequences, which has been successfully deployed to a news recommendation service. \cite{zhou2019deep} proposes a novel deep interest evolution network (DIEN) to model users' interest evolving process from behavior sequences. The key component is a new GRU structure enhanced with attention updating gates.  

However, a single vector-based recommender systems do not have enough expressive power to model a user, especially when the behavior sequence is long or multiple interests exist in the sequence. \cite{liu2019hi} proposes Hi-Fi Ark, which is a new user representation framework to comprehensively summarize user behavior history into multiple vectors. Similarly, \cite{li2019multi,10.1145/3397271.3401088} design a multi-interest extractor layer with various dynamic routing mechanisms to extract user's diverse interests. But neither of these models is designed for sequential recommender systems. \cite{wang2019modeling} proposes mixture-channel purpose routing networks (MCPRN) to detect the possible purposes of a user within a shopping session, thus it can recommend corresponding diverse items to satisfy a user's different purposes. \cite{ren2019lifelong} argues that existing RNN based models are only capable of dealing with relatively recent user behaviors. So it proposes a hierarchical RNN  with multiple update periods to better model user's lifelong sequential behaviors.  Recently, researchers find that Neural Turing Machines (NTM) \cite{DBLP:journals/corr/GravesWD14} is a very promising architecture to model a user's behavior sequence in a fine-grained level, so they study how to leverage this architecture for user modeling \cite{pi2019practice,chen2018sequential}. Our work is most related to RUM \cite{chen2018sequential}, we point out the difference between our work and RUM in Section \ref{sec:ourmodel} and reported experimental comparisons in Section \ref{sec:exp}.
\section{Conclusions}
A user's long behavior sequences usually include dynamic and diverse interests. Traditional sequential user modeling methods represent a user with a single vector, which is insufficient to describe the complicated and varied interests.  We propose a novel Sequential User Matrix (SUM) model, which leverages multiple channels to capture a users' multiple interests during user modeling. Our proposed SUM has three new components, including an interest-level and instance-level attention mechanism, a local proximity debuff mechanism, and a highway channel compared with existing sequential user models with memory networks. We conduct comprehensive experiments on two real-world datasets. The results demonstrate that our proposed model outperforms state-of-the-art methods consistently.

\bibliographystyle{ACM-Reference-Format}
\bibliography{sample-base}

\newpage
\appendix
\section{Appendix}

\subsection{Dataset Details}
\label{sec:dataset_appendix}
Here we describe the complete setting of datasets.

\textbf{Display Ads Dataset}. The SUM model is originally designed to support our online display advertising business in Bing Native Ads. We collect two weeks' ads clicking logs as data samples, and collect users' web behavior history prior to their corresponding ad click behavior for user modeling. The data samples are split into 70\%/15\%/15\% as training/validation/test dataset by users to avoid information leakage caused by repeated user behaviors. For more efficient offline modeling training, the user behavior sequences are truncated to 100. Some basic data statistics are reported in Table \ref{tab:dataset_stat}. For each positive instance (which is an ad click behavior from one user), we sample 1 negative instance from non-click impression, and randomly sample 3 negative instances by item popularity. 
Each web browsing behavior is represented by its web page title, e.g., \textsl{``Why Are People Rushing To Get This Stylish New SmartWatch? The Health Benefits Are Incredible''}. We use the CDSSM \cite{shen2014latent} model as a text-encoder to turn the raw text into a 128-dimension embedding vector. The embedding vector is then used as the static feature for a user page view behavior.

\textbf{Taobao Dataset}. This is a public e-commerce dataset\footnote{\url{https://tianchi.aliyun.com/dataset/dataDetail?dataId=649&userId=1}} collected from Taobao's recommender system. The original dataset contains several types of user behaviors such as page view and purchase. To make our two experimental datasets coherent, for Taobao Dataset, we take the purchase behaviors as target activities (which corresponds to the ads clicking behavior in Display Ads Dataset) and use page view behaviors for user modeling data (which corresponds to the web browsing behavior in Display Ads Dataset). To stay focused on studying users who have long behavior sequence, we only include users with more than 20 page view behaviors. We sort the user page view behaviors according to their timestamp so that we can get the last K user behaviors prior to the user's purchase activity. The user behavior sequences are truncated to 100, which aligns with the Display Ads dataset's setting. Some basic data statistics are reported in Table \ref{tab:dataset_stat}. Since we don't have the non-click impression logs in this dataset, all the negative instances are randomly sampled according to item popularity. For each positive instance, we sample 4 negative instances.
Unlike the \textsl{Display Ads} dataset, we don't have text descriptions for items. So we use item id and category id as one-hot features to represent items. The data samples are split into 70\%/15\%/15\% as training/validation/test dataset by users to avoid information leakage caused by repeated user behaviors. To make the training process more efficient, we pretrained item embeddings with a word2vec algorithm \footnote{\url{https://radimrehurek.com/gensim/models/word2vec.html}} on the training dataset. All models will load the pretrained item embeddings for better warm starting. This setting also helps to get rid of the auxiliary loss in \cite{zhou2019deep,pi2019practice}.

\subsection{Hyper-Parameter Settings} 
\label{sec:parameters_appendix}
We use grid-search to find the best hyper-parameters for each model on the validation set, then report the corresponding metrics on the test set. Experiments are repeated 3 times for each method and we take the best result in order to avoid being stuck in bad locally optimal solutions. The exploration range is: learning rate: \{0.0001, 0.0005, 0.001, 0.005, 0.01\}, L2 regularization weight for model parameters and embedding parameters: \{0.0, 0.0001, 0.001, 0.01\}, number of slots for NTM / RUM / MCPRN /  MIMN / HPMN / SUM and number of layers for SGRU: \{2, 3, 4, 5, 8, 10, 20\}, but if not explicitly mentioned, all models' performance are reported with the slot number of 5. For HRNN, we try the session-break time period in \{30, 60, 1440\} minutes. The update period $t$ is set to 2 according to \cite{ren2019lifelong} and the number of slots indicates the number of hierarchical layers. Batch size is fixed to 256 for all models. Embedding size of items are fixed on 128 on the Display Ads dataset; for the Taobao dataset, item ID embedding size is 64 and category ID embedding size is 16.  We concatenate item ID embedding and category ID embedding to represent an item. User states sizes are 128 for Ads dataset and 64 for Taobao dataset. The optimizer is \textsl{Adam}. A suggested configuration which works for most of the cases is that: learning rate is 0.0005, $\lambda$ is 0.0, number of slots/stacked layers is 5, session-break time period for HRNN is 1440.

\end{document}